# Kinematical Uniqueness of Loop Quantum Gravity

Christian Fleischhack

**Abstract.** We review uniqueness results for the kinematical part of loop quantum gravity. After sketching the general loop formalism, the holonomy-flux and the Weyl algebras are introduced. In both cases, then, diffeomorphism invariant representations are described.

## 1. Introduction

This Blaubeuren workshop has been dedicated to one of the greatest mysteries of modern physics – the unification of gravity and quantum theory. As can be seen from this edited volume, there are several different attempts to disclose at least glimpses of the merged theory. In general, there are two distinct strategies: either some radically new ideas are presented to formulate quantum gravity from scratch, or one focuses on fundamental results inside existing approaches. In this review, we will concentrate on the second issue. We will less discuss the general problem of quantum gravity itself, but study how far mathematical consistency may lead us to deeper insights into the conceptual foundations of one of the major possible routes to quantum gravity – loop quantum gravity.

The main achievement of loop quantization is to quantize gravity as it is – geometry. No additional structures are involved. In some sense, it is a minimalistic quantization. On the other hand, it does not include other interactions in nature. It may, of course, be questioned whether a quantum theory of gravity, or better a quantum theory of general relativity has to contain all existing forces. Indeed, classical gravity itself can be seen as some sort of derived interaction as it is determined intrinsically and purely geometrically by the principle of general relativity. There are approaches, first and foremost string theory, that imitate some sort of such an emergence strategy. However, certain features of gravity – in particular, its geometric origin – are usually quite hidden there. Loop quantum gravity tries to keep track of them. First of all, it admits a rigorous and nonperturbative treatment of diffeomorphism invariance being one of the most important and most peculiar implications of general relativity.

The overall conceptual framework, which loop quantum gravity provides, is reviewed in the article by Ashtekar in this volume. In the present article, we are going to focus on a special conceptual point: the current status of diffeomorphism invariance. We will see that this symmetry to a large extent uniquely fixes the kinematical framework of loop quantization. Note, however, that the theorems below are not to be seen as ultimate general statements on some kinematical structure of quantum gravity. They are applicable under the assumption only, that instead of metric and extrinsic curvature, parallel transports of Ashtekar connections and densitized dreibein fields are the canonical variables of gravity to be quantized. It is not the purpose of the present paper to decide whether these are indeed the appropriate classical variables to be quantized. Instead, we will discuss how unique the quantization of general relativity is if formulated in these terms.

The paper is organized as follows: First, Ashtekar's formulation of gravity and the loop variables used there are recalled. Then we study the configuration space with its compactification and calculate the basic Poisson brackets. In the main part of this article, we review the fundamental uniqueness results for diffeomorphism invariant representations of the holonomy-flux ∗-algebra and the Weyl algebra. We conclude with a discussion.

## 2. Ashtekar Variables

Using Ashtekar variables [1], canonical pure gravity can be written in terms of an $SU(2)$ connection $A_a^i$ and a densitized dreibein field $E_j^b$ fulfilling three types of constraints: Gauß, diffeomorphism and Hamilton constraint. The standard ADM formulation can be re-obtained by canonical transformations and symplectic reduction of the system w.r.t. the constraint surface given by the Gauß constraint. As for other gauge theories, this constraint generates the gauge transformations, here in some $SU(2)$ bundle over a Cauchy surface. The remaining constraints encode the invariance of general relativity w.r.t. diffeomorphisms of space-time. The so-called diffeomorphism[1] constraint generates diffeomorphisms within the Cauchy surface, while the Hamilton constraint governs the "dynamics" of the theory. Whereas the latter one is only very poorly understood to date, the Gauß and diffeomorphism constraints are well implemented as will be reviewed in this paper.

## 3. Loop Variables

The very first step to canonically quantize a classical system is to choose some appropriate set of classical variables to be quantized. This is a highly nontrivial task, since already this selection very much restricts possible quantizations. If too many classical variables get quantized, van Howe arguments may show that there is no nontrivial quantization at all. If, on the other hand, too little variables are in the game, information about certain physical degrees of freedom may be lost.

---
[1] More precisely, one should speak about the spatial diffeomorphism constraint.

### 3.1. Parallel Transports

If aiming at a functional integral quantization, one has to construct measures on the configuration space. Since we are dealing with Ashtekar's canonical gravity in the present case, we have to study gauge field theories. There, the configuration space is the space $\mathcal{A}/\mathcal{G}$ of (smooth) connections modulo gauge transformations. This space, however, is mathematically quite delicate: it is non-affine, non-compact, not a manifold and by far not finite-dimensional, whence measure theory is very complicated. As we will see, this problem can be solved using nonlocal parallel transport variables instead of connections themselves. The parallel-transport description is not only equivalent to that using connections, parallel transports even have a much nicer transformation behaviour w.r.t. bundle automorphisms, in particular, gauge transformations. In fact, instead of (locally)

$$A \longmapsto g^{-1}Ag + g^{-1}\mathrm{d}g$$

we have

$$h_A(\gamma) \longmapsto g_{\gamma(0)}^{-1}\, h_A(\gamma)\, g_{\gamma(1)} \qquad (1)$$

for all connections $A$, all sufficiently smooth paths $\gamma$ in the Cauchy slice and all gauge transformations $g$. Here, $h_A(\gamma) \equiv h_\gamma(A)$ denotes the parallel transport w.r.t. $A$ along $\gamma$. These variables now comprise the "position" variables to be quantized.[2]

### 3.2. Fluxes

The generalized momenta in Ashtekar gravity are densitized dreibein fields. In order to have non-singular classical Poisson brackets with parallel transports being smeared along one-dimensional objects, one has to smear the dreibein fields along one-codimensional objects, i.e. along hypersurfaces. More precisely, one first uses the antisymmetric symbol to turn the $E$ into a $\mathfrak{g}^*$-valued $(d-1)$-form $E'$ on some hypersurface $S$ and then smears it. The resulting flux variables $E_{S,f} := \int_S E'_i f^i$ with always $S$ being some hypersurface and $f : S \longrightarrow \mathfrak{g}$ some function, are the "momentum" variables selected for quantization.

## 4. Configuration Space

Any classical system is determined by the Poisson brackets between its basic variables. To quantize the system, first classical variables turn into abstract operators and Poisson brackets into commutators (or anti-commutators). Different quantizations then correspond to different representations of the algebras formed by (appropriate functions of) the abstract operators subject to the commutation relations. In our case, these algebras are the holonomy-flux $*$-algebra $\mathfrak{a}$ [15] and the

---

[2]In the gauge invariant case, only closed paths $\gamma$ are to be considered. These loops gave loop quantum gravity its name.

Weyl algebra $\mathfrak{A}$ [12], to be described below. In the former case, the parallel transports and the fluxes themselves generate the algebra; in the latter one, the fluxes are exponentiated. We will see that diffeomorphism invariance selects (to a large extent) unique representations of these algebras. Hence, quantization in the loop framework is unique at the kinematical level.

Before getting to the representation theory part, we have to study the Poisson relations and at least some parts of the algebras. They naturally lead, in particular, to a completion of the classical configuration space. Since, in the following, the results do not require the Ashtekar connection to be $\mathfrak{su}(2)$-valued and the Cauchy slice to be three-dimensional, we will deal with the general case of a pure gauge field theory over a principal fibre bundle $P(M, \mathbf{G})$ with $M$ being some manifold and $\mathbf{G}$ being some structure Lie group. We will only assume that $M$ is at least two-dimensional and that $\mathbf{G}$ is connected and compact. Now the configuration space consists of all smooth connections (modulo gauge transforms[3]) in $P$.

### 4.1. Semianalytic Structures

As we will see below, the basic Poisson bracket $\{(h_\gamma)^m_n, E_{S,f}\}$ with $m$ and $n$ being matrix indices, is completely determined by the intersection behaviour between the geometric ingredients $\gamma$ and $S$. This is no surprise, since diffeomorphism invariance reduces the theory essentially to topology. The intersection between graphs and hypersurfaces, however, may be quite wild in general. Already in the smooth category, there may exist accumulation points of intersections dividing paths into infinitely many parts. The only natural way out of this seems to always use real-analytic objects for paths, hypersurfaces and diffeomorphisms. However, there are two important bars. First, the intersection of analytic hypersurfaces, as arising for Poisson brackets between fluxes, is far from being analytic again. And second –and most importantly–, the use of analytic objects only, is very unphysical. In fact, gravity is a local theory. Already Einstein's hole argument uses the fact that gravity is invariant w.r.t. diffeomorphisms that are nontrivial only on some small compact neighbourhood in $M$. Analyticity, on the other hand, is very nonlocal: If you modify an analytic object, you modify it globally. In other words, we have to choose some other smoothness category, reconciling locality and analyticity. Fortunately, this is possible: The semianalytic category [14, 16, 6] provides us with the framework we need. Sloppily, there we are working with piecewise analytic objects. Analytic submanifolds are replaced by semianalytic sets, which themselves may have some kinks or creases, but at the same time consist of locally finitely many analytic submanifolds. For example, just imagine a sheet of a newspaper: It consist of three "analytic" parts, i.e., the two two-dimensional sheets with the columns and one one-dimensional fold. Unless the newspaper's fold is straightened, the full sheet is only semianalytic instead of analytic.

In what follows, we will always assume that we are working with semianalytic objects.

---

[3]For simplicity, by "configuration space" we will mean the space of connections without factorization.

### 4.2. Cylindrical Functions

Cylindrical functions, in general, are functions on the configuration space depending only on a finite number of values of the basic variables. Using here parallel transports rather than connections themselves, the algebra Cyl of cylindrical functions is now generated by all matrix elements $(T_{\phi,\gamma})^m_n := \phi(h_\gamma)^m_n$ of parallel transports $h_\gamma$, where $\gamma$ runs over all paths in $M$, $\phi$ runs over all (equivalence classes of) irreducible representations of $\mathbf{G}$, and $m$ and $n$ over all the corresponding matrix indices. Finite products of $(T_{\phi,\gamma})^m_n$ are called spin network functions [5][4] provided the underlying paths are distinct and form a graph. They already span Cyl as a vector space.

### 4.3. Generalized Connections

As we will see, the cylindrical functions form a subalgebra of both the holonomy-flux $*$-algebra $\mathfrak{a}$ and the Weyl algebra $\mathfrak{A}$. Therefore, representations of these algebras are automatically representations of Cyl. Since, on the other hand, Cyl is unital abelian, they are completely characterized by the Gelfand-Naimark theorem. In fact, denote the spectrum of the completion of Cyl by $\overline{\mathcal{A}}$. Then each representation of Cyl is the direct sum of representations of Cyl by multiplication operators on $L_2(\overline{\mathcal{A}}, \mu)$ for certain measures[5] $\mu$ on $\overline{\mathcal{A}}$. Of course, $\overline{\mathcal{A}}$ is compact and Hausdorff, and $\overline{\text{Cyl}}$ is isomorphic to the continuous functions on $\overline{\mathcal{A}}$. Moreover, by separation properties of parallel transports, $\mathcal{A}$ is even densely embedded into $\overline{\mathcal{A}}$. The elements of $\overline{\mathcal{A}}$ are, therefore, called generalized (or distributional) connections. W.l.o.g., we may consider $\overline{\mathcal{A}}$ as configuration space instead of $\mathcal{A}$. [2]

### 4.4. Projective Limit

Equivalently, $\overline{\mathcal{A}}$ may be described using projective limits. [3] For this, observe

$$\overline{\mathcal{A}} \;\cong\; \text{Hom}(\mathcal{P}, \mathbf{G})$$

where $\mathcal{P}$ denotes the groupoid[6] of paths in $M$. Indeed, each $h \in \text{Hom}(\mathcal{P}, \mathbf{G})$ defines a multiplicative functional $h_{\overline{A}}$ on $\overline{\text{Cyl}}$ via $h_{\overline{A}}\big((T_{\phi,\gamma})^m_n\big) := \phi(h(\gamma))^m_n$, which implies $h_{\overline{A}} \in \overline{\mathcal{A}}$. Now, each finite graph $\boldsymbol{\gamma}$ in $M$ defines a continuous projection

$$\pi_{\boldsymbol{\gamma}} : \overline{\mathcal{A}} \longrightarrow \text{Hom}(\mathcal{P}_{\boldsymbol{\gamma}}, \mathbf{G}) \cong \mathbf{G}^n$$

with $n := \#\boldsymbol{\gamma}$ via

$$\pi_{\boldsymbol{\gamma}}(h_{\overline{A}}) \;:=\; h_{\overline{A}}|_{\mathcal{P}_{\boldsymbol{\gamma}}} \;\widehat{=}\; h_{\overline{A}}(\boldsymbol{\gamma}) \;\equiv\; \big(h_{\overline{A}}(\gamma_1), \ldots, h_{\overline{A}}(\gamma_n)\big),$$

where $\mathcal{P}_{\boldsymbol{\gamma}}$ denotes the paths in $\boldsymbol{\gamma}$. Note that the edges $\gamma_1, \ldots, \gamma_n$ of $\boldsymbol{\gamma}$ generate $\mathcal{P}_{\boldsymbol{\gamma}}$ freely. Using the natural subgraph relation and defining

$$\pi^{\boldsymbol{\delta}}_{\boldsymbol{\gamma}} : \text{Hom}(\mathcal{P}_{\boldsymbol{\delta}}, \mathbf{G}) \longrightarrow \text{Hom}(\mathcal{P}_{\boldsymbol{\gamma}}, \mathbf{G})$$

---

[4] Note that, in the quoted reference, spin network functions have been defined a slightly different way in order to implement gauge invariance.

[5] From now on, all measures are assumed to be normalized, regular, and Borel.

[6] The groupoid structure is induced by the standard concatenation of paths modulo reparametrization and deletion/insertion of immediate retracings.

for $\boldsymbol{\gamma} \leq \boldsymbol{\delta}$ again by restriction, we get a projective system over the set of all finite graphs in $M$, whose projective limit $\varprojlim_{\boldsymbol{\gamma}} \operatorname{Hom}(\mathcal{P}_{\boldsymbol{\gamma}}, \mathbf{G})$ is $\overline{\mathcal{A}}$ again. We remark finally that for every $\psi \in \mathrm{Cyl}$ there is a graph $\boldsymbol{\gamma}$ in $M$ and some continuous $\psi_{\boldsymbol{\gamma}}$ with $\psi = \psi_{\boldsymbol{\gamma}} \circ \pi_{\boldsymbol{\gamma}}$.

### 4.5. Ashtekar-Lewandowski Measure

The main mathematical advantage of $\overline{\mathcal{A}}$ is its compactness. This opens the door to many structures from measure theory on, in particular, projective limits. In fact, if any sequence of measures $\mu_{\boldsymbol{\gamma}}$ on $\operatorname{Hom}(\mathcal{P}_{\boldsymbol{\gamma}}, \mathbf{G}) \cong \mathbf{G}^{\#\boldsymbol{\gamma}}$ is given with $\boldsymbol{\gamma}$ running over all graphs, such that the compatibility conditions

$$\mu_{\boldsymbol{\gamma}} \;\;=\;\; (\pi^{\boldsymbol{\delta}}_{\boldsymbol{\gamma}})_* \, \mu_{\boldsymbol{\delta}}$$

are fulfilled for all $\boldsymbol{\gamma} \leq \boldsymbol{\delta}$, then there is a unique measure $\mu$ on $\overline{\mathcal{A}}$ with $(\pi_{\boldsymbol{\gamma}})_* \mu = \mu_{\boldsymbol{\gamma}}$ for all $\boldsymbol{\gamma}$. This way, it is rather easy to define such measures. The most obvious choice provides us with the Ashtekar-Lewandowski measure $\mu_0$ [4]. One simply demands that each measure $\mu_{\boldsymbol{\gamma}}$ equals the Haar measure. Due to the Peter-Weyl theorem, $\mu_0$ is the only measure on $\overline{\mathcal{A}}$, where all nontrivial spin network functions have zero integral. The relevance of $\mu_0$ will become clear below.

### 4.6. Gauge Transforms and Diffeomorphisms

Since $\overline{\mathcal{A}}$ equals $\operatorname{Hom}(\mathcal{P}, \mathbf{G})$, we can naturally extend the action of gauge transforms from $\mathcal{A}$ to $\overline{\mathcal{A}}$. Simply generalize (1). [3] Diffeomorphisms can be implemented equally easily. The action of diffeomorphisms on $M$ is naturally lifted to the set of paths and graphs, whence to $\overline{\mathcal{A}}$ and $C(\overline{\mathcal{A}})$ as well. For instance, for any cylindrical function $\psi_{\boldsymbol{\gamma}} \circ \pi_{\boldsymbol{\gamma}}$ over some graph $\boldsymbol{\gamma}$, we have $\alpha_{\varphi}(\psi_{\boldsymbol{\gamma}} \circ \pi_{\boldsymbol{\gamma}}) = \psi_{\boldsymbol{\gamma}} \circ \pi_{\varphi(\boldsymbol{\gamma})}$. Here, $\alpha_{\varphi}$ denotes the action of the diffeomorphism $\varphi$ on Cyl. Note that both gauge transformations and diffeomorphisms leave the Ashtekar-Lewandowski measure on $\overline{\mathcal{A}}$ invariant.

## 5. Poisson Brackets

By means of the Poisson bracket, the fluxes may be regarded as a derivation [15] on the algebra Cyl of cylindrical functions:

$$X_{S,f}\, \psi \;\; := \;\; \{\psi, E_{S,f}\}. \tag{2}$$

In particular, the result is always a cylindrical function and depends only on the intersection behaviour between the hypersurface $S$ and the graph $\boldsymbol{\gamma}$ underlying the cylindrical function $\psi$. More precisely, if the transversal intersections of $\boldsymbol{\gamma}$ and $S$ are always vertices in $\boldsymbol{\gamma}$, then $X_{S,f}(\psi_{\boldsymbol{\gamma}} \circ \pi_{\boldsymbol{\gamma}})$ is again a cylindrical function over $\boldsymbol{\gamma}$, essentially given by left and right Lie derivatives on $\psi_{\boldsymbol{\gamma}}$. The explicit formula is given most easily using the Weyl operators being the exponentiated derivations.

### 5.1. Weyl Operators

Let $S$ be some oriented semianalytic subset in $M$, and let $d : S \longrightarrow \mathbf{G}$ be some function. The intersection properties of $S$ with paths $\gamma$ are encoded in some function $\sigma_S(\gamma)$. For its definition, first observe that, due to semianalyticity, each path can be decomposed into a finite number of paths whose interior is either fully contained in $S$ (internal path) or disjoint to $S$ (external path). If $\gamma$ is external path starting non-tangent at $S$, then $\sigma_S(\gamma)$ is $+1$ (or, resp., $-1$), if $\gamma$ starts to above (or, resp., below) $S$. Of course, "above" and "below" refer to the orientation of $S$. In any other case, we have $\sigma_S(\gamma) = 0$. Now, one checks very easily [9] that there is a unique map $\Theta_{S,d} : \overline{\mathcal{A}} \longrightarrow \overline{\mathcal{A}}$, with

$$h_{\Theta_{S,d}(\overline{A})}(\gamma) \;\; = \;\; d(\gamma(0))^{\sigma_S(\gamma)} \, h_{\overline{A}}(\gamma) \, d(\gamma(1))^{-\sigma_S(\gamma^{-1})} \tag{3}$$

for all external or internal $\gamma \in \mathcal{P}$ and all $\overline{A} \in \overline{\mathcal{A}}$.

Since $\Theta_{S,d}$ is a homeomorphism, its pull-back $w_{S,d}$ is an isometry on $C(\overline{\mathcal{A}})$. By the translation invariance of Haar measures, $\Theta_{S,d}$ even preserves the Ashtekar-Lewandowski measure $\mu_0$, turning $w_{S,d}$ into a unitary operator on $L_2(\overline{\mathcal{A}}, \mu_0)$. The operators $w_{S,d}$ are called Weyl operators. [12]

### 5.2. Flux Derivations

Now, it is straightforward to write down the Poisson bracket (2) more explicitly:

$$X_{S,f}\, \psi \;\; = \;\; \frac{\mathrm{d}}{\mathrm{d}t}\bigg|_{t=0} \psi \circ \Theta_{S,\mathrm{e}^{tf}} \, .$$

Note that $X_{S,f}$ maps Cyl to Cyl, although for its description again $\overline{\mathcal{A}}$ instead of $\mathcal{A}$ has been used and $\Theta_{S,d}$ generally fails to preserve $\mathcal{A}$. Note, moreover, that $X_{S,f}$ is linear in $f$. [15]

### 5.3. Higher Codimensions

Note that we did not restrict ourselves to the case of genuine hypersurfaces $S$, i.e., subsets of codimension 1. Although the Poisson bracket is originally given just for this case, the extension to higher codimensions can be justified easily. In fact, observe that the Weyl operator of the disjoint union of hypersurfaces equals the product of the (mutually commuting) Weyl operators of the single hypersurfaces. This way, e.g., the Weyl operator for an equator can either be defined directly as above or obtained from the Weyl operator of the full sphere times the inverses of the Weyl operators corresponding to the upper and lower hemisphere. Both options give the same result. [12]

## 6. Holonomy-Flux ∗-Algebra

As we have seen above, the fluxes can be regarded as a derivation on the algebra Cyl of cylindrical functions. The cylindrical functions, on the other hand, are in a natural way multiplication operators on Cyl. Therefore, both the position and the momentum variables may be seen as operators on Cyl.

### 6.1. Definition

Consider the vector space [15]

$$\mathfrak{a}_{\text{class}} := \text{Cyl} \times \Gamma_1^0(\overline{\mathcal{A}})$$

where the vector space $\Gamma_1^0(\overline{\mathcal{A}})$ of complexified generalized vector fields is generated by all the $X_{S,f} : \text{Cyl} \longrightarrow \text{Cyl}$ and given a Cyl-module structure. Moreover, $\mathfrak{a}_{\text{class}}$ is equipped with a Lie algebra structure by

$$\{(\psi_1, Y_1), (\psi_2, Y_2)\} = -(Y_1\psi_2 - Y_2\psi_1, [Y_1, Y_2]).$$

The quantum **holonomy-flux** $*$**-algebra** $\mathfrak{a}$ [15] is the $*$-algebra of all words in $\mathfrak{a}_{\text{class}}$ with concatenation as multiplication $\cdot$ and factorized by the canonical commutation relations

$$a \cdot b - b \cdot a = \mathrm{i}\,\{a, b\}, \tag{4}$$

induced by the Poisson brackets, and by the Cyl-module relations

$$\psi \cdot c + c \cdot \psi = 2\psi\,c.$$

Here, $a, b, c \in \mathfrak{a}_{\text{class}}$ and $\psi \in \text{Cyl}$. We indicate the corresponding equivalence classes by a hat. It turns out that the relations above do not impose additional relations on Cyl, whence Cyl is embedded into $\mathfrak{a}$. Therefore, we may drop the hats there. Moreover, note that $\mathfrak{a}$ is generated by all (equivalence classes corresponding to the) products of a cylindrical function with any, possibly vanishing number of flux derivations. The flux derivations are invariant w.r.t. the involution $*$; for cylindrical functions, we have $\psi^* = \overline{\psi}$. Finally, in a natural way, gauge transforms and diffeomorphisms act covariantly and by automorphisms on $\mathfrak{a}$.

### 6.2. Symmetric State

The representation theory of $\mathfrak{a}$, responsible for the superselection theory of loop quantum gravity, can be reduced via GNS to the study of states $\omega$ on $\mathfrak{a}$. Of particular interest are states that are symmetric w.r.t. certain algebra automorphisms $\alpha$, i.e., we have $\omega = \omega \circ \alpha$ for all such $\alpha$.

There is a state $\omega_0$ on $\mathfrak{a}$ which is invariant w.r.t. all bundle automorphisms on $P$, i.e., it is invariant w.r.t. all gauge transformations and all diffeomorphisms. It is given [15] by

$$\omega_0(a \cdot \widehat{Y}) = 0$$

and

$$\omega_0(\psi) = \int_{\overline{\mathcal{A}}} \psi \, \mathrm{d}\mu_0$$

for all $a \in \mathfrak{a}$, $Y \in \Gamma_1^0(\overline{\mathcal{A}})$, and $\psi \in \text{Cyl}$. The invariance heavily relies on the gauge and diffeomorphism invariance of the Ashtekar-Lewandowski measure.

As proven by Lewandowski, Okołów, Sahlmann and Thiemann, for some reasonable technical assumptions, $\omega_0$ is even *the only one* state on $\mathfrak{a}$, that is invariant w.r.t. all bundle automorphisms on $P$. In other words, there is only one possible quantization implementing gauge and diffeomorphism invariance. The proof of this theorem is outlined in the next subsection; the first two steps are similar to [15].

### 6.3. Uniqueness Proof

Let $\omega$ be some state on $\mathfrak{a}$ with the desired invariance properties. Assume further that all smearing functions $f$ are semianalytic and have compact support.

First, we prove $\omega(\widehat{X}^*\widehat{X}) = 0$ for all flux operators $X = X_{S,f}$. Using trivializations, partitions of unity, and linearity in $f$, we may restrict ourselves to the case that $S$ is given by the intersection of some cube around the origin in $\mathbb{R}^n$ with some linear hyperspace and that $f = f_\| \tau$ with supp $f_\| \subseteq S$ and $\tau \in \mathfrak{g}$. Consider now the sesquilinear form

$$(f_1, f_2) := \omega\big(\widehat{X}^*_{S,f_1\tau} \widehat{X}_{S,f_2\tau}\big)$$

for $f_1, f_2$ having supports as above. Choose some function $f_\perp$ on the perpendicular to $S$ through the origin with $f_\perp(0) = 1$, such that $\chi := f_\| \otimes f_\perp$ has support in the cube. Fix some vector $\vec{e}$ in $S$ and define

$$\varphi_\lambda := \mathrm{id} + \lambda \chi \vec{e}.$$

If $\lambda \in \mathbb{R}$ is sufficiently small, then $\varphi_\lambda$ is a diffeomorphism on $M$ preserving $S$ and being the identity outside the cube above. Now one immediately checks, that every function $F$ on $M$ with $F(\vec{x}) = \vec{e} \cdot \vec{x}$ on supp $\chi$ fulfills

$$\varphi_\lambda^* F = F + \lambda f_\|$$

on $S$, whence we get

$$(F, F) = (\varphi_\lambda^* F, \varphi_\lambda^* F) = (F, F) + 2\lambda \,\mathrm{Re}\, (F, f_\|) + \lambda^2 \,(f_\|, f_\|)$$

for all small $\lambda$ by diffeomorphism invariance. This implies, as desired, $(f_\|, f_\|) = 0$.

Second, the GNS construction for $\omega$ yields a Hilbert space $\mathfrak{H} := \overline{\mathfrak{a}/\mathfrak{i}}$ with scalar product $\langle [a], [b] \rangle = \omega(a^*b)$ and a representation $\pi_\omega$ of $\mathfrak{a}$ on $\mathfrak{H}$ with $\pi_\omega(a)[b] = [ab]$ for $a, b \in \mathfrak{a}$. Here, $\mathfrak{i}$ is the left ideal given by the elements $a \in \mathfrak{a}$ with $\omega(a^*a) = 0$. As just seen, all flux derivations are contained in $\mathfrak{i}$. Hence, $\omega$ vanishes on all products of cylindrical functions with one or more flux derivations, as $\omega_0$ does.

Finally, we show that $\omega$ equals $\omega_0$ for the remaining generators of $\mathfrak{a}$—all spin network functions. By the $*$-invariance of flux derivations and by relation (4), we have for all cylindrical functions $\psi$ and all flux derivations $X$

$$\omega(X\psi) = -\mathrm{i}\,\omega(\widehat{X} \cdot \psi) = -\mathrm{i}\,\overline{\omega(\overline{\psi} \cdot \widehat{X})} = 0.$$

Let us denote the homeomorphism on $\overline{\mathcal{A}}$ associated to $X$ by $\Theta_t$. By inspection, we see that $\psi \circ \Theta_t$ is always a finite linear combination of cylindrical functions, whereas the $t$-dependence (and differentiability information) is completely contained in the coefficients. Therefore, we may exchange $\omega$ and differentiation to get for all $t_0$

$$\tfrac{\mathrm{d}}{\mathrm{d}t}\big|_{t=t_0} \omega(\psi \circ \Theta_t) = \omega\big(X(\psi \circ \Theta_{t_0})\big) = 0.$$

Hence, $\omega(\psi \circ \Theta_t) = \omega(\psi)$ for all $t$. Let now $\psi$ be a nontrivial spin network function. We may decompose it into $(T_{\phi,\gamma})^m_n T$ with some edge $\gamma$ and some nontrivial $\phi$, where $T$ is a (possibly trivial) spin network function. Moreover, choose some hypersurface $S$ dividing $\gamma$ transversally at some point $x$ into $\gamma_1$ and $\gamma_2$, without intersecting any other edge in the graph underlying $\psi$. Finally, for each $g \in \mathbf{G}$ choose some smearing function $f$, such that $\mathrm{e}^{2f(x)} = g$. Directly from definition (3), we get (for the appropriate orientation of $S$)

$$\psi \circ \Theta_{S,\mathrm{e}^f} \;=\; \frac{\phi(g)^r_s}{\dim \phi}\, (T_{\phi,\gamma_1})^m_r (T_{\phi,\gamma_2})^s_n\, T.$$

Now, the nontriviality of $\phi$ gives the desired equation:

$$\begin{aligned}
\omega(\psi) &\;=\; \int_{\mathbf{G}} \omega(\psi)\,\mathrm{d}\mu_{\mathrm{Haar}}(g) \;=\; \int_{\mathbf{G}} \omega(\psi \circ \Theta_{S,\mathrm{e}^f})\,\mathrm{d}\mu_{\mathrm{Haar}}(g) \\
&\;=\; \omega\big((T_{\phi,\gamma_1})^m_r (T_{\phi,\gamma_2})^s_n\, T\big) \int_{\mathbf{G}} \frac{\phi(g)^r_s}{\dim \phi}\,\mathrm{d}\mu_{\mathrm{Haar}}(g) \;=\; 0 \;=\; \omega_0(\psi).
\end{aligned}$$

The proof concludes with $\omega(\mathbf{1}) = 1 = \omega_0(\mathbf{1})$.

## 7. Weyl Algebra

Roughly speaking, the holonomy-flux $*$-algebra contains exponentiated positions (parallel transports), but non-exponentiated momenta (fluxes). Working with unitary Weyl operators instead of self-adjoint flux derivations being their generators, allows us to study both positions and momenta in their exponentiated versions. Together they form the Weyl algebra of loop quantum gravity. This is similar to the Weyl algebra [8, 17], studied by Stone and von Neumann in quantum mechanics.

### 7.1. Definition

Recall that the cylindrical functions are bounded multiplication operators and the Weyl operators unitary operators on $\mathfrak{H}_0 := L_2(\overline{\mathcal{A}}, \mu_0)$. There, the diffeomorphisms act by unitaries as well.

Now, the $C^*$-subalgebra $\mathfrak{A}$ of $\mathcal{B}(\mathfrak{H}_0)$, generated by cylindrical functions and Weyl operators for constant smearing functions, is called **Weyl algebra** [12]. Its natural representation on $\mathfrak{H}_0$ will be denoted by $\pi_0$. Sometimes, we will consider the $C^*$-subalgebra $\mathfrak{A}_{\mathrm{Diff}}$ of $\mathcal{B}(\mathfrak{H}_0)$, generated by the Weyl algebra $\mathfrak{A}$ and the diffeomorphism group $\mathcal{D}$. One immediately sees that $\mathcal{D}$ acts covariantly on $\mathfrak{A}$.

### 7.2. Irreducibility

It is easy to see that $\mathfrak{A}$ is irreducible [12, 10, 11]. Let $f \in \mathfrak{A}'$. First, by $C(\overline{\mathcal{A}}) \subseteq \mathfrak{A}$, we have $\mathfrak{A}' \subseteq C(\overline{\mathcal{A}})' = L_\infty(\overline{\mathcal{A}}, \mu_0)$. Second, by unitarity of Weyl operators $w$, we have $f = w^* \circ f \circ w = w^*(f)$. Therefore, $\langle T, f \rangle = \langle T, w^*(f) \rangle = \langle w(T), f \rangle$ for every spin network function $T$. Each nontrivial $T$ may be decomposed into $T = (T_{\phi,\gamma})^m_n\, T'$ with some edge $\gamma$ and nontrivial $\phi$, where $T'$ is a (possibly trivial) spin network function. There are two cases: Either $\phi$ is abelian or nonabelian.

If $\phi$ is abelian, choose some hypersurface $S$ intersecting $\gamma$, but no edge underlying $T'$. Then[7] $w_{S,g}(T) = \phi(g^2)\, T$ for all $g \in \mathbf{G}$, whence

$$\langle T, f \rangle \;=\; \langle w_{S,g}(T), f \rangle \;=\; \overline{\phi(g^2)}\,\langle T, f \rangle.$$

Since $\phi$ is nontrivial, there is some $g \in \mathbf{G}$ with $\phi(g^2) \neq 1$. Hence, $\langle T, f \rangle = 0$.

If $\phi$ is nonabelian, then $\operatorname{tr}\phi$ has a zero [13]. Since square roots exist in any compact connected Lie group, there is a $g \in \mathbf{G}$ with $\operatorname{tr}\phi(g^2) = 0$. Choose now infinitely many mutually disjoint surfaces $S_i$ intersecting $\gamma$, but no edge used for $T'$. A straightforward calculation yields for $i \neq j$

$$\langle w_{S_i,g}(T), w_{S_j,g}(T) \rangle \;=\; \left|\frac{\operatorname{tr}\phi(g^2)}{\dim \phi}\right|^2 \;=\; 0.$$

Now, $\langle w_{S_i,g}(T), f \rangle = \langle T, f \rangle = \langle w_{S_j,g}(T), f \rangle$ implies $\langle T, f \rangle = 0$ again.

Altogether, $\langle T, f \rangle = 0$ for all nontrivial spin network functions $T$, whence $f$ is constant. Consequently, $\mathfrak{A}'$ consists of scalars only.

### 7.3. Diffeomorphism Invariant Representation

Beyond irreducibility, the natural representation $\pi_0$ of $\mathfrak{A}$ has some special properties. First, it is diffeomorphism invariant, i.e., there is a diffeomorphism invariant vector in $\mathfrak{H}_0$ (the constant function) and the diffeomorphisms act covariantly on $\mathfrak{A}$. Second, this vector is even cyclic. Third, $\pi_0$ is regular, i.e., it is weakly continuous w.r.t. the Weyl operator smearings $g$. Now, as in the case of the holonomy-flux $*$-algebra, these properties already (to a large extent) distinguish $\pi_0$ among the $C^*$-algebra representations of $\mathfrak{A}$. [12]

### 7.4. Uniqueness Proof

Let $\pi : \mathfrak{A} \longrightarrow \mathcal{B}(\mathfrak{H})$ be some regular representation of $\mathfrak{A}$ on some Hilbert space $\mathfrak{H}$. Assume $\pi$ diffeomorphism invariant, i.e., $\pi$ is the restriction of some representation $\pi_{\mathrm{Diff}}$ of $\mathfrak{A}_{\mathrm{Diff}}$ on $\mathfrak{H}$ having some diffeomorphism invariant vector. Moreover, let this vector be cyclic for $\pi$. Technically, let us assume that the dimension of $M$ is at least three and that all the hypersurfaces used for the definition of Weyl operators are "reasonably" triangulizable. Finally, let the diffeomorphisms act naturally (as to be explained below). We are going to sketch the proof [12, 11] that $\pi$ equals $\pi_0$.

First, observe that $C(\overline{\mathcal{A}})$, by the denseness of cylindrical functions, is contained in $\mathfrak{A}$. Now, by the general theory of $C^*$-algebras, the restriction of $\pi$ to $C(\overline{\mathcal{A}})$ is the direct sum of canonical representations $\pi_\nu$ of $C(\overline{\mathcal{A}})$ by multiplication operators on $L_2(\overline{\mathcal{A}}, \mu_\nu)$ for some measures $\mu_\nu$. The constants $\mathbf{1}_\nu$ are cyclic for $\pi_\nu$. We may choose some $\mathbf{1}_c$ to be diffeomorphism invariant and cyclic for $\pi$.

Second, for simplicity, let us assume $\mathbf{G}$ abelian. Fix $\varepsilon > 0$. Let $\psi$ be some nontrivial spin network function, i.e., $\psi = (h_\gamma)^n\, T$ for some spin network function $T$ and some $n \neq 0$. Assume $\langle \mathbf{1}_c, \pi(\psi)\mathbf{1}_c\rangle_{\mathfrak{H}} \neq 0$. Now, define for some "cubic" hypersurface $S$

$$w_t := w^S_{\mathrm{e}^{\mathrm{i}t/2}} \quad \text{and} \quad v_t := \frac{1}{2^m} \sum_{k=1}^{2^m} \alpha_{\varphi_k}(w_t).$$

---

[7] We shortly write $w_{S,g}$ instead of $w_{S,d}$ if $d(x)$ equals $g \in \mathbf{G}$ everywhere on $S$.

Here, each $\varphi_k$ is a diffeomorphism winding $\gamma$, such that it has exactly $m$ punctures with $S$. Each $k$ corresponds to a sequence of $m$ signs $+$ or $-$ denoting the relative orientations of $S$ and $\varphi_k(\gamma)$ at the $m$ punctures. (The windings are the reason for $M$ to be at least three-dimensional.) Then

$$v_t(\psi) \;=\; \left(\frac{\mathrm{e}^{\mathrm{i}nt} + \mathrm{e}^{-\mathrm{i}nt}}{2}\right)^m \cdot \psi$$

and

$$\|(v_t - \mathrm{e}^{-\frac{1}{2}m(nt)^2})\,\psi\|_\infty \;\leq\; O(m(nt)^4)\,\|\psi\|_\infty.$$

Hence, for any small $t$, there is some $m = m(nt, \varepsilon)$ with

$$\begin{aligned}
\varepsilon &< O(m(nt)^2)\,|\langle \mathbf{1}_c, \pi(\psi)\mathbf{1}_c\rangle_{\mathfrak{H}}| - O(m(nt)^4)\,\|\psi\|_\infty \\
&\leq |(1 - \mathrm{e}^{-\frac{1}{2}m(nt)^2})\langle \mathbf{1}_c, \pi(\psi)\mathbf{1}_c\rangle_{\mathfrak{H}}| - |\langle \mathbf{1}_c, \pi[(v_t - \mathrm{e}^{-\frac{1}{2}m(nt)^2})\psi]\mathbf{1}_c\rangle_{\mathfrak{H}}| \\
&\leq |\langle \mathbf{1}_c, \pi[(v_t - \mathbf{1})\psi]\mathbf{1}_c\rangle_{\mathfrak{H}}|.
\end{aligned}$$

Now, for each small $t$, there is a diffeomorphism $\varphi$ with

$$\begin{aligned}
\varepsilon &\leq |\langle \mathbf{1}_c, \pi[(w_t - \mathbf{1})(\alpha_\varphi(\psi))]\mathbf{1}_c\rangle_{\mathfrak{H}}| \\
&\leq 2\,\|\mathbf{1}_c\|_{\mathfrak{H}}\,\|(\pi(w_t) - \mathbf{1})\mathbf{1}_c\|_{\mathfrak{H}}\,\|\pi(\alpha_\varphi(\psi))\|_{\mathcal{B}(\mathfrak{H})} \\
&= 2\,\|\mathbf{1}_c\|_{\mathfrak{H}}\,\|(\pi(w_t) - \mathbf{1})\mathbf{1}_c\|_{\mathfrak{H}}\,\|\psi\|_\infty,
\end{aligned}$$

by diffeomorphism invariance. The final term, however, does not depend on $\varphi$, whence by regularity it goes to zero for $t \to 0$ giving a contradiction. Consequently, $\langle\psi\rangle_{\mu_c} \equiv \langle \mathbf{1}_c, \pi(\psi)\mathbf{1}_c\rangle_{\mathfrak{H}} = 0$ implying $\mu_c = \mu_0$ or, for simplicity, $c = 0$.

Finally, let $w$ be a Weyl operator assigned to some ball or simplex $S$, possibly having higher codimension. Observe that, for $w$ commuting with $\alpha_\varphi$, we have

$$\langle \pi(\alpha_\varphi(\psi))\mathbf{1}_0, \pi(w)\mathbf{1}_0\rangle_{\mathfrak{H}} \;=\; \langle \pi(\psi)\mathbf{1}_0, \pi(w)\mathbf{1}_0\rangle_{\mathfrak{H}}.$$

for all spin-network functions $\psi$. Next, for nontrivial $\psi$, we choose infinitely many diffeomorphisms $\varphi_i$ that leave $S$ invariant, but move the respective graph underlying $\psi$ to mutually distinct ones. Then each $\varphi_i$ commutes with $w$ and we have

$$\delta_{ij} \;=\; \langle \alpha_{\varphi_i}(\psi), \alpha_{\varphi_j}(\psi)\rangle_{\mathfrak{H}_0} \;\equiv\; \langle \pi(\alpha_{\varphi_i}(\psi))\mathbf{1}_0, \pi(\alpha_{\varphi_j}(\psi))\mathbf{1}_0\rangle_{\mathfrak{H}}.$$

This is possible, unless the ball or simplex has dimension 1 or 2. In fact, there the graph underlying $\psi$ may coincide with $S$ or its boundary. In these cases, the argumentation is technically more involved; here, we only refer to [12]. In the other cases, we now have, with $P_0$ being the canonical projection from $\mathfrak{H}$ to $\mathfrak{H}_c \cong \mathfrak{H}_0$,

$$0 \;=\; \langle \pi(\psi)\mathbf{1}_0, \pi(w)\mathbf{1}_0\rangle_{\mathfrak{H}} \;=\; \langle \psi, P_0\pi(w)\mathbf{1}_0\rangle_{\mathfrak{H}_0}.$$

Thus, $P_0\pi(w)\mathbf{1}_0 = c(w)\,\mathbf{1}_0$ with $c(w) \in \mathbb{C}$, whence also $(\mathbf{1} - P_0)\pi(w)\mathbf{1}_0$ generates $L_2(\overline{\mathcal{A}}, \mu_0)$. The naturality[8] of $\pi$ w.r.t. the action of diffeomorphisms implies that $\pi(w)\mathbf{1}_0$ is diffeoinvariant itself. Since $S$ is assumed to be a ball or simplex (with lower dimension than $M$), there is a semianalytic diffeomorphism mapping $S$ to itself, but inverting its orientation. Now, we have

$$\pi(w)^2 \mathbf{1}_0 \;=\; \pi(w)\pi(\alpha_\varphi)\pi(w)^*\pi(\alpha_\varphi)^*\mathbf{1}_0 \;=\; \mathbf{1}_0$$

implying $\pi(w)\mathbf{1}_0 = \mathbf{1}_0$ by taking the square root of the smearing. The proof is completed using triangulizability (disjoint unions of semianalytic subsets correspond to products of Weyl operators) and cyclicity.

## 8. Conclusions

Let us compare the two main results reviewed above.

### 8.1. Theorem – Self-Adjoint Case

Let $\mathfrak{a}$ be given as in Subsection 6.1, whereas only those flux derivations are used that correspond to oriented, one-codimensional, semianalytic $C^k$ hypersurfaces $S$ and to compactly supported, semianalytic $C^k$ smearing functions $f$ on $S$, with some $k > 0$. Moreover, let $M$ be at least two-dimensional and let $\mathbf{G}$ be connected, compact and nontrivial. Then, $\omega_0$ is the only state on $\mathfrak{a}$ that is invariant w.r.t. all semianalytic bundle transformations, acting covariantly on $\mathfrak{a}$.

### 8.2. Theorem – Unitary Case

Let $\mathfrak{A}$ be given as in Subsection 7.1, whereas only those Weyl operators are used that correspond to oriented, widely[9] triangulizable, at least one-codimensional, semianalytic $C^0$ subsets $S$ and to constant smearing functions $d$ on $S$. Moreover, let $M$ be at least three-dimensional and let $\mathbf{G}$ be connected, compact and nontrivial. Then $\pi_0$ is the only regular representation of $\mathfrak{A}$ having a cyclic and diffeomorphism invariant vector, whereas the semianalytic $C^0$ diffeomorphisms act naturally and covariantly on $\mathfrak{A}$.

### 8.3. Comparison

Mathematically, both theorems look quite related. In fact, for instance, in the case of Lie theory in finite dimensions, the representations of a compact Lie group are always determined by the representations of the corresponding Lie algebra. At most, it may happen that some of the algebra representations do not extend to group representations by some global "discrete" restrictions. Here, the situation seems similar. In the Lie case, the self-adjoint generators of the Lie group unitaries form the Lie algebra; here, the self-adjoint flux derivations in the holonomy-flux ∗-algebra $\mathfrak{a}$ are the generators of Weyl operators in the Weyl algebra $\mathfrak{A}$. Hence, the uniqueness result for $\mathfrak{a}$ should imply a uniqueness results for $\mathfrak{A}$. However, appearances are deceiving. The main point is the hugeness of the algebras in the game, whence on the self-adjoint level domain issues are to be taken seriously. In the Lie case, it can be proven that they are waived; here, however, we have to deal with infinite-dimensional objects, even nonseparable Hilbert spaces, whence general results are quite scarce. Indeed, by now, there is no direct mathematical relation between these two uniqueness results known.

---

[8] A representation $\pi$ is called natural w.r.t. the action of diffeomorphisms iff, for each decomposition of $\pi|_{C(\overline{\mathcal{A}})}$ into cyclic components $\pi_\nu$, the diffeomorphism invariance of the $C(\overline{\mathcal{A}})$-cyclic vector for $\pi_{\nu_1}$ implies that for $\pi_{\nu_2}$, provided the measures $\mu_1$ and $\mu_2$ underlying these representations coincide.

[9] A triangulation $(K, \chi)$ is called wide iff for every $\sigma \in K$ there is some open chart in $M$ containing the closure of $\chi(\sigma)$ and mapping it to a simplex in that chart.

Moreover, there are quite some other differences between the two settings. For instance, in the holonomy-flux case, the smearings are always compactly supported, whereas in the Weyl case they are constant. The former idea allowed us to use the linearity of the flux derivations w.r.t. the smearing functions; the latter one opened the road to use triangulations into geometrically simpler objects to get rid of the non-linearities of the smearings. It is not known how far either theorems remain valid in the other cases.

A disadvantage of the Weyl case is to need at least three dimensions. The main advantage of the Weyl case, of course, is to circumvent all domain problems ubiquitous in the holonomy-flux case. The only remnant is the regularity of the Weyl operators themselves. However, this only requires that every Weyl operator has a self-adjoint generator with some dense domain; in the holonomy-flux case, these domains all have to coincide. On the other hand, the naturality of the action of diffeomorphisms is only required in the Weyl case, whereas the scope of this assumption is not known yet. Whether this might be cancelled by assuming not only diffeomorphism invariance, but also gauge invariance as in the holonomy-flux case is not known.

### 8.4. Discussion

Physically, nevertheless, both results are far-reaching. Up to the technical issues mentioned in the previous paragraphs, they show that there is essentially only one quantization of Ashtekar gravity within the loop formalism. Therefore, for this framework, they approach the relevance of the Stone-von Neumann theorem for quantum mechanics. This celebrated result, proven some 75 years ago, is responsible for the (to a large extent) uniqueness of quantization of classical mechanics.

In fact, let us consider one-dimensional classical mechanics. For quantization, it is assumed that the position and momentum variables $x$ and $p$ turn into self-adjoint operators that fulfill $[\widehat{x},\widehat{p}] = \mathrm{i}$ as induced by the Poisson brackets. Now, these operators generate weakly continuous one-parameter subgroups of unitaries: $U(\sigma) := \mathrm{e}^{\mathrm{i}\sigma\widehat{x}}$ and $V(\lambda) := \mathrm{e}^{\mathrm{i}\lambda\widehat{p}}$. The commutation relation above turns into

$$U(\sigma)V(\lambda) \;\;=\;\; \mathrm{e}^{\mathrm{i}\sigma\lambda}\,V(\lambda)U(\sigma). \qquad (5)$$

The Stone-von Neumann theorem now tells us the following [17]: Each pair $(U,V)$ of unitary representations of $\mathbb{R}$ on some Hilbert space that satisfies the commutation relations (5) for all $\sigma, \lambda \in \mathbb{R}$, is equivalent to multiples of the Schrödinger

representation

$$U(\sigma) = \mathrm{e}^{\mathrm{i}\sigma x}. \quad \text{and} \quad V(\lambda) = L^*_\lambda$$

by multiplication and pulled-back translation operators on $\mathbb{R}$. The desired uniqueness now follows from irreducibility. In other words, assuming continuity and irreducibility, all "pictures" of quantum mechanics are equivalent. They are physically indistinguishable.

Although the relevance of the loop quantum gravity theorems may indeed be related to the Stone-von Neumann theorem, mathematically there is a major difference between them. In the latter case, both the position and the momentum operators are assumed to be regular, i.e., weakly continuous. In the gravity case, however, the position operators are no longer subject to this requirement. Even more, they are *not* weakly continuous. In fact, only the parallel transports turn into well-defined quantum operators; the connections, in some sense their original generators, are ill defined at the quantum level. The continuity is lost, when the cylindrical functions have been used to form basic variables. Of course, since the continuity is lost already at the level of the algebra and not only at that of representations, this does not weaken the results reviewed in the present articles.

Incidently, when the regularity assumption is dropped in the case of quantum mechanics, other representations appear that are non-equivalent to the Schrödinger representation. One of them is given by almost-periodic functions, which lead to the Bohr compactification of the real line. The Hilbert space basis is given by $\{|x\rangle \mid x \in \mathbb{R}\}$, and the operators $U$ and $V$ act by

$$U(\sigma)|x\rangle = \mathrm{e}^{i\sigma x}|x\rangle \quad \text{and} \quad V(\lambda)|x\rangle = |x + \lambda\rangle.$$

Obviously, $V$ is not continuous. Hence, $\widehat{p}$ is not defined, but the operator $V(\lambda)$ corresponding to $\mathrm{e}^{\mathrm{i}\lambda p}$ only. This type of representation fits much more the pattern we described above. It is very remarkable that just this Bohr-type representation reappears in loop quantum cosmology and leads to a resolution of the big bang singularity [7].

## 9. Acknowledgements

The author is very grateful to Bertfried Fauser, Jürgen Tolksdorf and Eberhard Zeidler for the kind invitation to the Blaubeuren workshop on quantum gravity.

Christian Fleischhack
Max-Planck-Institut für Mathematik in den Naturwissenschaften
Inselstraße 22–26
04103 Leipzig
Germany                                                        February 12, 2006

*Present Address (May 17, 2015):*
Institut für Mathematik
Universität Paderborn
33095 Paderborn
Germany
e-mail: `fleischh@math.upb.de`